\documentclass[aps,prd,twocolumn,nofootinbib,amsmath,amssymb]{revtex4}
\usepackage{times}
\usepackage{graphicx}

\begin{document}

\title{Mass Relations for the Quark-Diquark Model}
\author{Sultan Catto$^{a,{\dag}}$}
\affiliation{Physics Department\\University Center and The Graduate School \\ The City University of New York \\ 365 Fifth Avenue, New York, NY 10016-4309 \\ and \\ Center for Theoretical Physics\\ The Rockefeller University \\ New York, New York 10021-6399}
\begin{abstract}
Quark model with potentials derived from QCD, including the quark-diquark model for excited hadrons gives mass formulae in very good agreement with experiment and goes a long way in explaining the approximate symmetries and supersymmetries of the hadronic spectrum, including the symmetry breaking mechanism.
\end{abstract}
\maketitle

\section{Introduction}
There is a good phenomenological evidence that in a rotationally excited baryon a quark-diquark ($q-D$) structure is favored over a three quark ($qqq$) structure. Regge trajectories for mesons and baryons are closely parallel; both have a slope of about $0.9 GeV^{-2}$. At large spin two of the quarks form a diquark ($D=qq$), a bilocal object at one end of a bag, the remaining quark being at the other. For the light quarks it was shown ([1]-[3])  that the underlying quark-diquark symmetry leads to supersymmetric $SU(6/21)$ symmetry between mesons and baryons. This new scheme uses split octonion algebra that renders the algebraic description of color degrees of freedom, supresses color-symmetric and space-symmetric quark configurations, and leads to existence of exotic meson (diquark-antidiquark) states for which there is now a good deal of experimental evidence. In this note we show derivation of hadronic mass formulae and show they are in remarkable agreement with experiments. A new and  extended semirelativistic and relativistic mass formulae for hadrons, based on supersymmetry schemes that make use of algebra of octonions (largest of Hurwitz' division algebras) are considered in detail in other publications([4]-[7]).

\section{Mass Formulae}
The ground state energy eigenvalue $E$ of the Hamiltonian                              
can be estimated by using the Heisenberg uncertainty principle. This leads                              
to the replacement of $r$ by $\Delta r$ and $p_{r}$ by 
                             
\begin{equation}                              
\Delta p_{r} = \frac{1}{2} (\Delta r)^{-1}, ~~~~~~(h = 1).                              
\end{equation}                              
                              
Then $E$ as a function of $\Delta r$ is minimized for the value of                              
$r_{0}$ of $\Delta r$. The $r_{0}$ corresponds to the Bohr radius for                              
the bound state. The confining energy associated with this Bohr radius                               
is obtained from the linear confining potential $S(r) = br$, so that the effective masses of the constituents become                              

\begin{equation}                              
M_{1} = m_{1} + \frac{1}{2} S_{0}, ~~~~                              
M_{2} = m_{2} + \frac{1}{2} S_{0}, ~~~~(S_{0} = b r_{0})
\end{equation}                                                            
For a meson $m_{1}$ and $m_{2}$ are the current quark masses while                              
$M_{1}$ and $M_{2}$ can be interpreted as the constituent quark                                
masses. Note that even in the case of vanishing quark masses associated                              
with perfect chiral symmetry, confinement results in non zero                              
constituent masses that spontaneously break the $SU(2) \times SU(2)$                               
symmetry of the $u$, $d$ quarks.                              
                              
Let us illustrate this method on the simplified spin free Hamiltonian                              
involving only the scalar potential. In the center of mass system,                                
$\mbox{\boldmath $p$}^{(1)}+\mbox{\boldmath $p$}^{(2)}=0$, or                                
$\mbox{\boldmath $p$}^{(1)}=-\mbox{\boldmath $p$}^{(2)}=\mbox{\boldmath $p$}$.                                
The semi-relativistic hamiltonian of the system is then given by                                

\begin{equation}                                
E_{12}~~ \Phi=\sum_{i=1}^{2}\sqrt{(m_{i}+\frac{1}{2}br)^{2}+                              
\mbox{\boldmath $p$}^{2}}~~\Phi~.                                
\end{equation}                                
Taking $m_{1}=m_{2}=m$ for the quark-antiquark system, we have                                

\begin{equation}                                
E_{12}~~\Phi= 2\sqrt{(m+\frac{1}{2}br)^{2}+p^{2}_{r}+                              
\frac{\ell(\ell+1)}{r^{2}}} ~~\Phi           \label{eq:e}                                
\end{equation}                                
where we have written the momentum part in spherical coordinates.                              
                              
Putting                              
\begin{equation}                              
b=\mu^{2},~~~~~\rho = \mu ~~r,                              
\end{equation}                              
for the $q-\bar{q}$ system we find $E_{12}$ by minimizing the function                              

\begin{equation}                                
E_{q \bar{q}}~~= 2\sqrt{(m+\frac{1}{2} \mu \rho)^{2}+                              
\frac{\mu^{2}}{\rho^{2}} (\ell+\frac{1}{2})^{2}} ~.                                
\end{equation}                                
                                
For $u$ and $d$ quarks, $m$ is small and can be neglected so that                                

\begin{equation}                              
E^{2} = \mu^{2} [\rho^{2} + \rho^{-2} (2 \ell + 1)^{2}]                              
\end{equation}                              
which has a minimum for                              

\begin{equation}                              
\rho^{2} = \rho_{0}^{2} = 2 \ell + 1                              
\end{equation}                              
giving                              

\begin{equation}                              
E_{min}^{2} = E^{2} (\rho_{0}) = 4 \mu^{2} (\ell+\frac{1}{2})        \label{eq:min}       
\end{equation}                              
Thus, we obtain a linear Regge trajectory with                              

\begin{equation}                              
\alpha^{'} = \frac{1}{4} \mu^{-2} = \frac{b}{4}~.                       \label{eq:mina}       
\end{equation}                              
Also $ \mbox{\boldmath $J$}= \mbox{\boldmath $\ell$} +\mbox{\boldmath $S$},$ where $\mbox{\boldmath $S$}$ arises from the quark spins. Experimentally                              

\begin{equation}                              
\alpha^{'} = 0.88 (GeV)^{-2}                              
\end{equation}                              
for mesons giving the value $0.54$ GeV for $\mu$. A more accurate calculation (see [8]) gives

\begin{equation}                              
\alpha^{'} = (2 \pi \mu^{2})^{-1}, ~~~~~ \mu \sim 0.43 GeV~.   \label{eq:alp}                              
\end{equation}      
                              
The constituent quark mass can be defined in two ways                              

\begin{equation}                              
M_{c}(\ell) = \frac{1}{2} E_{min} = \mu \sqrt{\ell + \frac{1}{2}}       \label{eq:mc}                              
\end{equation}                              
or                              

\begin{equation}                              
m_{c}^{'}(\ell)=S_{0} = \frac{1}{2} \mu \rho_{0} =                               
\frac{\mu}{\sqrt{2}} \sqrt{\ell + \frac{1}{2}}~.                    \label{eq:mca}          
\end{equation}                              
The first definition gives for $\ell = 0$, 

\begin{equation}
M_{c} = 0.31 GeV ~~~for~~~\mu= 0.43    \label{eq:mcb}                              
\end{equation}                              
in the case of $u$ and $d$ quarks.                              
                              
When the Coulomb like terms are introduced in the simplified Hamiltonian                              
(\ref{eq:e}) with negligible quark masses one obtains                              

\begin{equation}                              
E=\frac{\mu}{\rho} [-\bar{\alpha}  + \sqrt{\rho^{4} +      
(2\ell + 1)^{2}}]                              
\end{equation}                              
with                              

\begin{equation}                              
\bar{\alpha} =\frac{4}{3} \alpha_{s} ~~~for (q\bar{q}),~~~                              
\bar{\alpha} =\frac{2}{3} \alpha_{s} ~~~for (q q)~.                              
\end{equation}                              
In the energy range around $1$ GeV, $\alpha_{s}$ is of order of unity.                              
Estimates range from $0.3$ to $3$. Minimization of $E$ gives                              

\begin{equation}                              
E_{0}=\mu u_{0}^{\frac{-1}{4}} (-\bar{\alpha} + \sqrt{u_{0} +(2\ell + 1)^{2}})                              
\end{equation}                              
where                              

\begin{equation}                              
u_{0}(\epsilon)=\rho_{0}^{4}= (2 \ell + 1)^{2} (1+\frac{1}{2} \beta^{2} +                              
\epsilon \sqrt{2} \beta                              
\sqrt{\ell+\frac{1}{8} \beta^{2}}),                              
\end{equation}                    
          
\begin{equation}                              
\epsilon = \pm{1},~~~~\beta=\frac{\bar{\alpha}}{(2 \ell + 1)}~.                              
\end{equation}                              
                              
The minimum $E_{0}$ is obtained for $\epsilon = -1$, giving to second                              
order in $\beta$:
                              
\begin{equation}                              
E_{0} = \mu \sqrt{2(2 \ell + 1 )} (1 - \frac{\beta}{\sqrt{2}} -                              
3 \frac{\beta^{2}}{8})~.       \label{eq:ez}                              
\end{equation}                              
                              
Linear Regge trajectories are obtained if $\beta^{2}$ is negligible.                              
Then for mesons        
                      
\begin{equation}                              
E_{0}^{2} = 4 \mu^{2} \ell + 2 \mu^{2} (1-\sqrt{2} \bar{\alpha})~.                              
\end{equation}                              
$\beta^{2}$ is negligible for small $\ell$ only if we take the lowest                              
estimate for $\alpha_{s}$, giving $0.4$ for $\bar{\alpha}$ in the $q \bar{q}$                              
case. For mesons with $u$, $d$ constituents, incorporating their spins through                              
the Breit term we obtain approximately                              

\begin{equation}                              
m_{\rho} \simeq m_{\omega} = E_{0} + \frac{c}{4},~~                                                             
m_{\pi}= E_{0} - \frac{3c}{4}, ~~c = K~~ \frac{\Delta V}{M_{q}^{2}}
\end{equation}
where $M_{q}$ is the constituent quark mass. This gives                              

\begin{equation}                              
E_{0}= \frac{(3 m_{\rho} + m_{\pi})}{4} = 0.61 GeV~.                               
\end{equation}                              
                              
The Regge slope being of the order of $1 GeV$ an average meson mass                              
of the same order is obtained from Eq.(\ref{eq:ez}) in the linear 
trajectory approximation. To this approximation $\bar{\alpha}$ should be treated like                               
a parameter rather than be placed by its value derived from QCD under                               
varying assumptions. Using Eq.(\ref{eq:alp}) for $\mu$ one gets a better                               
fit to the meson masses by taking $\alpha_{s} \sim 0.2$.                              

Turning now to baryon masses, we must first estimate the diquark mass. We                              
have for the $qq$ system                              

\begin{equation}                              
M_{D}= \mu (\sqrt{2} - \frac{2}{3} \alpha_{s})~,                \label{eq:dq}           
\end{equation}                              
that is slightly higher than the average meson mass                              

\begin{equation}                              
\tilde{m}= \mu (\sqrt{2} - \frac{4}{3} \alpha_{s})~.                           \label{eq:con}      
\end{equation}                              
                              
Here we note that $E$ is not very sensitive to the precise value of the QCD running coupling constant in the $GeV$ range. Taking $\alpha_{s} \sim 0.3$ changes $E^{qq}$ from $0.55$ to $0.56 GeV$.      
      
Note that Eq.(\ref{eq:dq}) gives $m_{D} = 0.55 GeV$. For excited $q-\bar{q}$ and $q-D$ systems if the rotational excitation energy is large compared with $\mu$, then both the $m_{D}$ and the Coulomb term $- \frac{4}{3} \frac{\alpha_{s}}{r}$ (same for $q-D$ and $q-\bar{q}$ systems) can be neglected. Thus, for both ($q-D$) [excited baryon] and $q-\bar{q}$ [excited meson] systems we have Eq.(\ref{eq:min}), namely      

\begin{equation}      
(E^{q-D})^{2} \sim (E^{q-\bar{q}})^{2} \sim 4 \mu^{2} \ell + 2 \mu^{2}      
\end{equation}      
giving again Eq.(\ref{eq:mina}), i.e.      

\begin{equation}      
(\alpha^{'})_{q-D} = (\alpha^{'})_{q-\bar{q}} \cong \frac{1}{4 \mu^{2}}~~~~{\rm or} ~~(\frac{1}{2 \pi \mu^{2}})      
\end{equation}      
as an explanation of hadronic supersymmetry in the nucleon and meson Regge spectra. We also have, extrapolating to small $\ell$:      

\begin{equation}      
\Delta (M^{2})^{q-D} = \Delta (m^{2})^{q-\bar{q}} = 4 \mu^{2} \Delta \ell = \frac{1}{\alpha^{'}} \Delta \ell ~.      
\end{equation}        
For $\Delta \ell = 1$ we find      

\begin{equation}      
m_{\Delta}^{2} - m_{N}^{2} = m_{\rho}^{2} - m_{\pi}^{2} ~.      
\end{equation}      
      
This relationship is same as the one proved in our earlier paper ([1]) through the assumption that $U(6/21)$ symmetry is broken by an operator that behaves like $s = 0$, $I = 0$ member of $35 \times 35$ representations of $SU(6)$, which is true to $5\%$. It corresponds to a confined quark approximation with $\alpha_{s} = 0$.      
      
The potential model gives a more accurate symmetry breaking ($\alpha_{s} \sim 0.2$):      
\begin{equation}      
\frac{9}{8} (m_{\rho}^{2} - m_{\pi}^{2}) = m_{\Delta}^{2} - m_{N}^{2}      
\end{equation}      
with an accuracy of $1\%$.       
      
This mass squared formula arises from the second order iteration of the $q-D$, $q-\bar{q}$ Dirac equation. The factor $\frac{9}{8}$ comes from      

\begin{equation}      
\frac{1}{2} (\frac{4}{3} \alpha_{s})^{2} = \frac{8}{9} \alpha_{s}^{2} ~.      
\end{equation}

At this point it is more instructive to derive a first order mass formula. Since the constituent quark mass $M_{q}$ is given by Eq.(\ref{eq:mc})                              
$(\ell = 0)$, we have                              

\begin{equation}                              
M_{q} = \frac{\mu}{\sqrt{2}}~,                              
\end{equation}                              
so that                              
\begin{equation}                              
\bar{m} = 2 M_{q} ( 1 -\frac{\sqrt{2}}{3} \alpha_{s}) \simeq 1.9 M_{q} .                              
\end{equation}                              
                              
When the baryon is regarded as a $q-D$ system, each constituent gains an                               
effective mass $\frac{1}{2} \mu \rho_{0}$ which was approximately the                              
effective mass of the quark in the meson. Hence, the effective masses of                              
$q$ and $D$ in the baryon are                               
\begin{equation}                              
m_{q}^{'} \simeq M_{q}, ~~~~~m_{D}^{'} = M_{D} + M_{q} \simeq 3 M_{q}.                              
\end{equation}                              
                              
The spin splittings for the nucleon $N$ and the $\Delta$ are given by                              
the Breit term                              

\begin{equation}                              
\Delta M = K \Delta V \frac{\mbox{\boldmath $S$}_{q}                                
\mbox{\boldmath $\cdot S$}_{D}}{m_{q}^{'} m_{D}^{'}}~.                              
\end{equation}                              
                              
For the nucleon with spin $\frac{1}{2}$ the term ${\bf S}_{q} \cdot {\bf S}_{D}$                              
gives $-1$ while it has the value $\frac{1}{2}$ for $\Delta$ with spin                              
$\frac{3}{2}$. Using the same $K$ for mesons and baryons which are                              
both considered to be a bound state of a color triplet with a color                               
antitriplet we can relate the baryon splitting $\Delta M$ to the meson                              
splitting $\Delta m$ for which ${\bf S}_{q} \cdot {\bf S}_{\bar{q}}$ takes the                              
values $\frac{1}{4}$ and $\frac{-3}{4}$. Hence we find                              

\begin{equation}                              
\Delta M = M_{\Delta} - M_{N} = \frac{3}{2}\cdot \frac{K \Delta V}{m_{q}^{'} m_{D}^{'}} = 
\frac{1}{2} \cdot \frac{K \Delta V}{ M_{q}^{2}}~,
\end{equation}
and

\begin{equation}
\Delta m = \frac{K \Delta V}{M_{q}^{2}}                              
\end{equation}                              
which leads to a linear mass formula                              

\begin{equation}                              
\Delta M = \frac{1}{2} \Delta m                              
\end{equation}                              
which is well satisfied, and has been verified before using the three                               
quark constituents for the baryon([3]).                              
                              
The formation of diquarks which behave like antiquarks as far as QCD is                               
concerned is crucial to hadronic supersymmetry and to quark dynamics for                               
excited hadrons. The splittings in the mass spectrum are well understood                              
on the basis of spin-dependent terms derived from QCD. This approach to                               
hadronic physics has led to many in depth investigations recently. For extensive references we refer to recent papers by Lichtenberg and collaborators ([9]) and by Klempt ([10]).
                
To see the symmetry breaking effect, note that the mass of a hadron will take the approximate form

\begin{equation}
m_{12}=m_1+m_2+K\frac{{\bf S}_1\cdot{\bf S}_2}{m_1m_2}
\end{equation}
where $m_i$ and ${\bf S}_i$ ($i=1,2$) are respectively the constituent mass and the spin of a quark or a diquark. The spin-dependent Breit term will split the masses of hadrons of different spin values. If we assume $m_q=m_{\bar q}=m$, where $m$ is the constituent mass of $u$ or $d$ quarks, and denote the mass of a diquark as $m_D$, then this approximation gives

\begin{equation}
m_\pi= (m_{q\bar{q}})_{s=0}=2m-K\frac{3}{4m^2}~,
\end{equation}

\begin{equation}
m_\rho= (m_{q\bar{q}})_{s=1}=2m+K\frac{1}{4m^2}~'
\end{equation} 

\begin{equation}
m_\Delta= (m_{qD})_{s=3/2}=m+m_D+K\frac{1}{2mm_D}~,
\end{equation}
and

\begin{equation}
m_N= (m_{qD})_{s=1/2}=m+m_D-K\frac{1}{mm_D}~.
\end{equation}

Eliminating $m$, $m_D$ and $K$, we obtain a mass relation 
\begin{equation}
\frac{8}{3}\cdot \frac{2m_{\Delta}+m_N}{3m_{\rho}+m_{\pi}}=1+\frac{3}{2} \cdot \frac{m_\rho-m_\pi}{m_\Delta-m_N}
\end{equation} 
which agrees with experiment to 13\%.

\vspace{0.8cm}
($a$) Invited talk presented at XXIV International Colloquium on Group Theoretical Methods in Physics (Paris, July 15-20, 2002). To be published in: {\em Physical and Mathematical Aspects of Symmetries.} Eds. J-P. Gazeau, R. Kerner, J-P. Antoine, S. M\'etens and J-Y. Thibon.  

\vspace{0.8cm}
$^\dag$ Work Supported by DOE grants DE-AC-0276-ER3074 and 3075, and PSC-CUNY Research Awards. 

\end{document}